\documentclass{jfm}
\usepackage{graphicx}
\usepackage{epstopdf, epsfig}
\usepackage{amsmath}
\usepackage[usenames]{color}
\usepackage[utf8]{inputenc}
\usepackage[dvipsnames]{xcolor}

\newcommand{\tp}[1]{{\textcolor{blue}{#1}}}

\newcommand{\bld}{\boldsymbol}
\newcommand{\balpha}{\mbox{\boldmath$\alpha$}}

\title{Swimming of microorganisms in quasi-2D membranes}

 \author{Carlos Alas\aff{1}, 
 Thomas~R.~Powers\aff{2}, and
  Tatiana Kuriabova\aff{1}
   \corresp{\email{tkuriabo@calpoly.edu}}
 }

\affiliation{\aff{1}Department of Physics, California Polytechnic State University, San Luis Obispo, California 93407, USA
\aff{2}School of Engineering and Department of Physics, Brown University, Providence, Rhode Island 02912, USA}

\begin{document}

\maketitle

\begin{abstract}
Biological swimmers frequently navigate in geometrically restricted media. We study the prescribed-stroke
problem of swimmers confined to a
planar viscous membrane embedded in a bulk fluid of different viscosity.   
In their motion, microscopic swimmers disturb the fluid in both the membrane and the bulk. The flows that emerge have a combination of two-dimensional (2D) and three-dimensional (3D) hydrodynamic features, and such flows are referred to as quasi-2D. The cross-over from 2D to 3D hydrodynamics in a quasi-2D fluid is controlled by the Saffman length, a length scale given by the ratio of 
the 2D membrane viscosity to  
the 3D viscosity of the embedding bulk fluid. We have developed a computational and theoretical approach based on the boundary element method and the Lorentz reciprocal theorem to study the swimming of microorganisms for a range of values of the Saffman length.  
We found that a flagellum propagating transverse sinusoidal waves in a quasi-2D membrane can develop a swimming speed exceeding that in pure 2D or 3D fluids, while the propulsion of a two-dimensional squirmer is slowed down by the presence of the bulk fluid. 
\end{abstract}


\section{Introduction}
Microscopic biological organisms have adapted to a viscous world in which their inertia is inconsequential to their locomotion. For a typical micro-scale organism the Reynolds number $\Rey  = \rho U L/\eta$ is small. For example, {\it Escherichia coli}  
has a characteristic length $L \sim10$\,$\mu$m and a characteristic swimming 
speed $U\sim10\,\mu$m/s  
in water (density $\rho \approx 10^3$\,$\rm{kg/m^3}$ and dynamic viscosity $\eta \approx 10^{-3}$\,Pa\,s),
leading to a negligibly small Reynolds number $\Rey  = \rho U L/\eta\sim 10^{-5}$--$10^{-4}$ \citep{Purcell1977}. 
At this scale, a swimmer reacts instantaneously to any forces, oblivious to any history of prior dynamics \citep{Purcell1977,  Childress1981, LaugaPowers2009}, and so 
the primary method of 
locomotion
for macroscopic swimmers 
such as fish and humans, which relies on trading momentum with the fluid to generate a propulsive force, is ineffective here. Rather, the net translations of a microorganism are determined by the sequence of configurations it adopts to swim, independent of its deformation rate. 
Microswimmers must continually paddle or deform their bodies in a swimming pattern with non-reciprocal forward and reverse strokes to manipulate the drag forces for propulsion \citep{Purcell1977, Childress1981, LaugaPowers2009}.  

It is common for microorganisms to swim in geometrically confined media: in channels, near surfaces and interfaces, and in films. A significant amount of theoretical and experimental work has been devoted to studying the effects of confinement on the motion of microscopic swimmers near solid walls \citep{PedleyKessler1987, Lauga2006, Berke2008, Drescher2009, LiTang2009, OrMurray2009, Or2011, CrowdyOr2010, Li2011, Spagnolie2012, Molaei2014, Ishimoto2016}, near fluid-fluid interfaces \citep{Guasto2010, DiLeonardo2011, Wang2013, LopezLauga2014, MasoudStone2014, StoneMasoud2015}, and in thin fluid layers atop a solid substrate \citep{Lambert2013, Mathijssen2016a, Mathijssen2016b, Ota2018}. 

\begin{figure}
\centerline{\includegraphics[width=6cm]{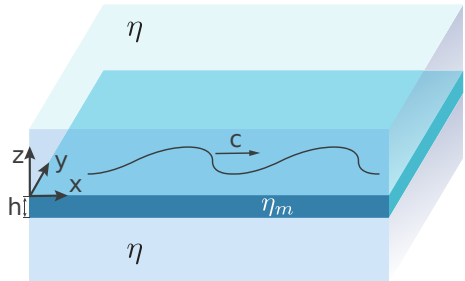}}
\caption{Illustration of a flagellum confined to a plane of a thin membrane of 2D viscosity $\eta_m$ sandwiched between two semi-infinite slabs of bulk fluid of 3D viscosity $\eta$. The flagellum propagates transverse planar waves traveling with speed $c$ with respect to the flagellum.} 
\label{fig:taylors-swimmer}
\end{figure}

Motivated by recent experimental and theoretical studies of bacteria swimming in biofilms, in freely-suspended thin films \citep{Aranson2007, Sokolov2007}, and on active proteins mimicking biological swimmers in lipid membranes \citep{Huang2012}, we study here the hydrodynamics of swimming microorganisms in a thin membrane. 
We treat the membrane as a continuous incompressible viscous fluid 
film of very small thickness. Flow fields in such a membrane are uniform throughout the thickness of the membrane. In contrast to a thin-film model that involves integration of three-dimensional (3D) hydrodynamic equations across the thickness of the film, our membrane model is intrinsically two-dimensional (2D), in the sense that the motion of molecules within the membrane in the direction normal to the plane of the membrane is forbidden. 
The membrane is embedded in a 3D fluid of different viscosity. The motion of a swimmer in the membrane generates flows both in the membrane and in the surrounding fluid (see figure~\ref{fig:taylors-swimmer}). While there have been some recent investigations of the hydrodynamics of swimmers in a thin layer of fluid sandwiched between fluids of a different viscosity~\citep{LeoniLiverpool2010,RowerPadidarAtzberger2019}, this problem is still largely unexplored.

As was demonstrated by \citet{SaffmanDelbruck1975} and \citet{Saffman1976}, the amount of momentum imparted by the membrane to the bulk fluid is controlled by a hydrodynamic length scale, the so-called Saffman length $\ell_S$, given by the ratio of the 2D membrane viscosity $\eta_m$ to the 3D viscosity of the bulk fluid $\eta$,  $\ell_S = \eta_m/(2\eta)$.
If the membrane is perturbed by a localized force (applied in the plane of the membrane) at a point $\boldsymbol{x}$,  for distances (measured from $\boldsymbol{x}$) much smaller than the Saffman length, the effect of the flows in the bulk on the membrane hydrodynamics is negligible. 
In this region the fluid velocity in the membrane decays slowly (logarithmically) with distance, as in purely 2D fluids. On the other hand, for distances much larger than the Saffman length, the contribution of the bulk fluid to the membrane dynamics is significant, and the membrane flow field decays inversely with the distance, a behavior consistent with 3D dynamics.

 \citet{Levine_MacKintosh2002} (LM) derived a Green function for a more general case of viscoelastic membranes.
 In the case of a purely viscous membrane that we consider here, there is no elastic response of the membrane, and a disturbance caused by a force results in the velocity field alone. The velocity of the membrane at position $\boldsymbol{x}^\prime$ due to an in-plane, localized force  $\boldsymbol{f}(\boldsymbol{x}) = \boldsymbol{f}\delta(\boldsymbol{x})$ is determined by the LM response tensor $\balpha({\bld x} - {\bld x}')$, 
\begin{equation}
\label{eq:response}
{\bld v}({\bld x}) = \frac{1}{4\pi\eta_m}  \balpha({\bld x} - {\bld x}')  \bcdot  {\bld f}({\bld x}').
\end{equation}
Here ${\bld x}$ and ${\bld x}'$ are in-plane vectors with components $(x, y)$ and $(x^\prime, y^\prime)$, respectively (refer also to figure~\ref{fig:taylors-swimmer} for our choice of the coordinate system). The response function $\balpha({\bld x} - {\bld x}')$ in equation~(\ref{eq:response}) plays the role of the Oseen tensor in $3$D hydrodynamics.

As was shown by LM, the response function may be split into `parallel' and `transverse' contributions. In the component form we have
\begin{equation}
\label{eq:response-functions}
\alpha_{\alpha\beta}({\bld x}) = \alpha_\|(|{\bld x}|) \hat{x}_\alpha \hat{x}_\beta + \alpha_\perp(|{\bld x}|)
[\delta_{\alpha\beta} - \hat{x}_\alpha\hat{x}_\beta],
\end{equation}
where  $\alpha,\beta=x,y$, and  
 $\hat{x}_{\alpha}$ is the $\alpha$ component of  
 the unit vector $\hat{{\bld x}} ={\bld x}/|{\bld x}|$. In our notation, $\alpha_{\alpha\beta}$  corresponds to  $-i\omega\alpha_{\alpha\beta}$ in the LM theory. 
The scalar functions $\alpha_\|$ and $\alpha_\perp$ are given by
\begin{eqnarray}
\label{eq:alpha-parallel}
\alpha_\|(\kappa)&=& \frac{\pi}{\kappa} {\bld H}_1(\kappa) - \frac{2}{\kappa^2} - \frac{\pi}{2}[Y_0(\kappa) + Y_2(\kappa)]  \\
\label{eq:alpha-perp}
\alpha_\perp(\kappa) &=&  \pi{\bld H}_0(\kappa) - \frac{\pi}{\kappa}{\bld H}_1(\kappa) + \frac{2}{\kappa^2} - \frac{\pi}{2}[Y_0(\kappa)- Y_2(\kappa)], \nonumber
\end{eqnarray}
where ${\bld H}_\nu$ are Struve functions and $Y_\nu$ are Bessel functions of the second kind \citep{Abramowitz1965}; $\kappa=|{\bld x}|/\ell_S$ is the non-dimensionalized distance between the point of application of the force and the point where the membrane velocity response is measured. Both $\alpha_\|(\kappa)$ and $\alpha_\perp(\kappa)$ diverge logarithmically as $\kappa\to 0$, while for large $\kappa$ we have $\alpha_\|(\kappa)\sim 1/\kappa$ and $\alpha_\perp(\kappa)\sim 1/\kappa^2.$

In the small Reynolds number regime the inertia term in the Navier-Stokes equation can be neglected. We assume that the membrane has thickness $h$ and choose a coordinate system with the origin at the top surface of the membrane with $z=0$ (therefore, the bottom side of the membrane is at $z=-h$). The dynamics of the membrane embedded in a bulk fluid is governed by a modified Stokes equation and the incompressibility condition \citep{Saffman1976},
\begin{eqnarray}
\label{eq:Stokes1}
-\bnabla p + \frac{\eta_m}{h} \bnabla^2 \boldsymbol{v}+ \frac{2\boldsymbol{f}}{h} = 0,  \qquad   \bnabla \cdot \boldsymbol{v}=0,
\end{eqnarray}
where $p$ and $\boldsymbol{v}$  are the pressure and velocity fields of the membrane. The flows in the membrane set the bulk fluid into motion. The resulting flows
 in the embedding fluid, in their turn, exert traction on the membrane. Since the membrane is  
 only a few molecular layers  thick, 
 the traction due to the bulk fluid produces a flow that is uniform throughout the thickness of the membrane, i.e. the fluid velocity does not depend on $z$ for $-h < z < 0$.  In equation~(\ref{eq:Stokes1}) the coupling between the membrane and the bulk fluid is described by the force per unit volume $2\boldsymbol{f}/h$, with 
\begin{equation}
\boldsymbol{f} = \eta \frac{\partial \boldsymbol{v}^{(3D)}}{\partial z}{\Big |}_{z=0}, 
\end{equation}
where $\boldsymbol{v}^{(3D)}$ is the bulk fluid velocity.  The factor of two in  equation~(\ref{eq:Stokes1}) is due to an equal force $\boldsymbol{f}$ acting on the bottom side of the membrane. Equation~(\ref{eq:Stokes1}) can be written in a more compact form that we will use later, 
\begin{equation}
\label{eq:Stokes2}
\bnabla \cdot \boldsymbol{\sigma} = - \frac{2 \boldsymbol{f}}{h},
\end{equation}
where $\boldsymbol{\sigma}$ is the stress field of the membrane.

Being inertialess and swimming in the absence of external influences, a swimmer must maintain a zero net force $\boldsymbol{F}(t)$ and a zero torque $\boldsymbol{L}(t)$ on its body at every time instant,
\begin{eqnarray}
\label{eq:net-force1}
\boldsymbol{F}(t) &=& \int_{S} \boldsymbol{\sigma}\cdot \boldsymbol{n} \, \mathrm{d} S,\\
\label{eq:net-torque}
 \boldsymbol{L}(t) &=& \int_{S} \boldsymbol{x} \times (\boldsymbol{\sigma}\cdot \boldsymbol{n})\, \mathrm{d} S,
\end{eqnarray}
where the integration is over the surface of the swimmer and $\boldsymbol{n}$ is a unit vector normal to the surface and pointing away from the swimmer. 

Many microorganisms such as spermatozoa, {\it E. Coli}, and {\it Caulobacter crescentus} swim by moving thin extensions (flagella) on their bodies. Some critters, like {\it Paramecium}, are covered in thousands of short hair-like appendages called cilia and propel themselves through a coordinated beating of these cilia. 
Since our primary goal is to study how the confinement to the plane of a membrane affects the swimming dynamics of a microorganism
(rather than a detailed study of a particular microorganism), we consider here only minimal theoretical models of flagellated and ciliated microorganisms.

In \S\ref{Sec:infinite-flagellum} we consider a headless, infinitely long `flagellum' of infinitesimally small thickness propagating planar sinusoidal waves along its body. This is a one-dimensional analog of the Taylor swimmer \citep{Taylor1951}, an infinite plane in viscous fluid passing transverse sinusoidal waves. We recover Taylor's result for the swimming velocity in the limiting case 
of a pure 2D hydrodynamics (the membrane in vacuum). We find that the membrane incompressibility condition imposes a constraint on the fluid dynamics that allows the flagellum to achieve much higher swimming speed than in pure 2D and 3D fluids for large ratios of the wavelength to the Saffman length.  

In \S\ref{sec:finite} we study the propulsion of a flagellum of finite length and find its swimming speed and efficiency. In  \S\ref{Sec:infinite-flagellum} and \S\ref{sec:finite} we apply the boundary-element method (BEM) that two of us have recently developed in work on hydrodynamic interaction of inclusions in freely-suspended smectic films \citep{Qi2014, Kuriabova2016, Qi2017}. 

In \S\ref{Sec:Lorentz} we formulate the Lorentz reciprocal theorem for a quasi-2D fluid and derive an equation for the swimming speed. We discuss the advantage of the method based on the Lorentz reciprocal theorem over the BEM for studying microscopic swimmers with swimming patterns that do not change the overall shape of the swimmers' bodies (for example, swimmers propagating longitudinal compressive waves along their bodies). As an example of such a critter, we consider a simple model of a two-dimensional `squirmer', a disk with a prescribed tangential velocity along its circumference. We find that unlike its flagellated counterpart, a squirmer does not benefit from the presence of the bulk fluid: its swimming speed is lower than that in a purely 2D fluid. 

In \S\ref{Sec:Conclusion} we discuss our results and suggest further directions of investigation. 

\section{Infinitely long flagellum in a quasi-2D membrane} 
\label{Sec:infinite-flagellum}

The Taylor swimmer \citep{Taylor1951} is an infinite  swimming 2D sheet in a 3D viscous fluid that propagates transverse waves of amplitude $b$ and wave speed $c = \omega/q$.  In a frame moving with the swimming sheet (co-moving frame) the 
shape of the sheet is described by    
\begin{equation}
\label{eq:modulation}
y = b \sin(q x - \omega t) = b\sin\xi,
\end{equation}
where $\xi = q x - \omega t$ denotes the wave phase. 

Taylor showed that the sheet with such a one-dimensional modulation travels, relative to the fluid at infinity, with a speed $U/c = (bq)^2/2 + O((bq)^4)$ in the direction opposite to the wave velocity. 
We consider here a one-dimensional analog of the Taylor swimmer: an infinitely long, infinitesimally thin flagellum confined to the plane of a viscous membrane embedded in bulk fluid (see figure~\ref{fig:taylors-swimmer}) with prescribed motion given by equation~(\ref{eq:modulation}), with $x$ and $y$ parametrizing the shape of the flagellum.

A swimming velocity of an infinitely long flagellum is time-independent. Indeed, two snapshots of the waving flagellum taken at the same point on the $x$-axis differ only by a
shift  $\Delta x$ along the $x$-axis. Thus, a temporal shift at a fixed point $x$ is equivalent to a spatial displacement along the flagellum. A swimmer moving as a whole has the same translational velocity along its entire length. The swimming velocity, being invariant with respect to translations along the $x$-axis must be invariant with respect to translations in time as well.
Therefore, we can calculate the swimming speed for a single time instant and set $t=0$ in equation~(\ref{eq:modulation}). 

As in \citep{Taylor1951} we consider the case of an inextensible flagellum.  In order to calculate the portion of the swimmer's velocity due to its distortion, we calculate the position of a material point of the flagellum as a function of time. In a frame moving with the wave (with speed $c$ relative to the co-moving frame) the shape of the flagellum does not change. The Cartesian coordinates for this frame are $(x',y)$, where $x'=x-ct$. In this reference frame a material particle of the flagellum travels a distance $\Lambda$ equal to the arclength of the flagellum spanned by one (linear) wavelength $\lambda$ during one period of oscillation $T = 2\pi/\omega$,
\begin{equation}
\label{eq:arc-Lambda}
\Lambda =  \frac{1}{q}\int_0^{2\pi} \sqrt{1+ (bq)^2 \cos^2 \xi} \mathrm{d} \xi.
\end{equation}
 We will call $\Lambda$ the arcwise wavelength. The material particle's speed is, therefore, 
 \begin{eqnarray}
 \label{eq:Q-speed}
 C &=& \frac{c}{2\pi} \int_0^{2\pi}  \sqrt{1+ (bq)^2 \cos^2 \xi} \mathrm{d} \xi\\
 &\approx&c\left(1+\frac{1}{4}b^2q^2-\frac{3}{64}b^4q^4\right).
 \end{eqnarray}
 To determine the position of a material particle of the flagelleum, we define the material coordinate $S$ to be the arclength coordinate $s$ of a material point at $t=0$. In the frame in which the nodes of the wave are fixed, arclength is related to the Cartesian coordinate $x'$ by
 \begin{eqnarray}
 s&=&\int_0^{x'}\sqrt{1+(bq)^2\cos^2(qx')}\mathrm{d}x'\\
 &\approx&x'+\frac{b^2q}{8}\left[2qx'+\sin(2qx')\right]-\frac{b^4q^3}{256}\left[12qx'+8\sin(2qx')+\sin(4qx')\right].
 \end{eqnarray}
 Reverting the series leads to
\begin{equation}
 x'\approx s-\frac{b^2q}{8}\left[2qs+\sin(2qs)\right]+\frac{b^4q^3}{256}\left[28qs+16qs\cos(2qs)+16\sin(2qs)+5\sin(4qs)\right].
 \end{equation}
 Using $y=b\sin qx'$ and $s=S-Ct$ leads to the position of the material point labeled by $S$ as a function of time $t$:
 \begin{eqnarray}
 x&\approx& S-\frac{b^2q}{8}\left[2q S+\sin2(qS-\omega t)\right]\nonumber\\
 &&+\frac{b^4q^3}{256}\left[28qS+16qS\cos2(qS-\omega t)+16\sin2(qS-\omega t)+5\sin4(qS-\omega t)\right],\\
 y&\approx&b\sin(q S-\omega t)\nonumber\\
 &&-\frac{b^3}{16}\left[4q^3S\cos(qS-\omega t)+q^2\sin(q S-\omega t)+q^2\sin3(qS-\omega t)\right].
 \end{eqnarray}
In the co-moving frame the components of a material particle's velocity $\bld{u}_S$ are given by
\begin{equation}
\bld{u}_S=\left(\left.\frac{\partial{x}}{\partial t}\right|_S,\left.\frac{\partial{y}}{\partial t}\right|_S\right).\label{uScomps}
\end{equation}
For an arbitrary value of $b$, the material particle's velocity can be calculated numerically using
\begin{eqnarray}
\label{eq:inextensible_ux}
u_{S, x} &=&- C\cos\theta_S  + c \\
\label{eq:inextensible_uy}
u_{S, y} &=&-C\sin\theta_S,
\end{eqnarray}
with $\displaystyle \tan\theta_S = \left.\frac{\partial y}{\partial x}\right|_S$.
The total velocity of a material particle relative to the fluid at infinity (for $y\to \pm \infty$) is the sum of the surface disturbance and swimming velocities, $\bld{u}_S + \bld{U}$.

The linearity of Stokes equations allows us to model the fluid velocity field  in the membrane as a superposition of fluid velocities due to a (yet unknown) force density $\bld{f}(\bld{x})$ along the flagellum:
\begin{equation}
\label{eq:membrane-velocity}
\bld{v}(\bld{x}) =  \frac{1}{4\pi\eta_m} \int_{\Gamma} \balpha(\bld{x} - \bld{x}^\prime)\bcdot \bld{f}(\bld{x}^\prime) \mathrm{d} \bld{x}^\prime,
\end{equation}
where the integration is along the (infinite) contour of the flagellum. 
 
The spatial periodicity of the flagellum modulation implies the invariance of the flow field and the force density $\bld{f}(\bld{x})$ in equation~(\ref{eq:membrane-velocity}) under translations along the $x$-axis by an integer multiple of the wavelength $\lambda=2\pi/q$. Defining $x_m= x  + m \lambda$, for all integers $m$,  
the integration on the RHS of equation~(\ref{eq:membrane-velocity}) can be reduced to 
integration over a single wavelength. 
We impose a no-slip boundary condition on the surface of the flagellum by setting the fluid velocity equal to the velocity of the material point on the flagellum, 
\begin{equation}
\label{eq:int-one-lambda}
\bld{u}_S(\bld {x}) + \bld{U} =  \frac{1}{4\pi\eta_m} \int_{\Gamma_0} \sum_{m=-\infty}^{\infty} \balpha(\bld{x} - \bld{x}_m^\prime)\bcdot \bld{f}(\bld{x}^\prime) \mathrm{d} \bld {x}^\prime,
\end{equation}
where $\bld{x}$ and $\bld{x}^\prime$ indicate the points on the flagellum that belong to a one-wavelength `window' $\Gamma_0$ and $\bld{x}_m^\prime = ( x^\prime + m \lambda,  y^\prime )$. The surface disturbance velocity $\bld{u}_S(\bld{x})$ on the LHS of equation~(\ref{eq:int-one-lambda}) is given by 
equation~(\ref{uScomps}). 

To close the system of equations for the force density $\bld{f}(\bld{x})$ and the swimming velocity $\bld{U}$, we also require the net force on the flagellum be equal to zero,
\begin{equation}
\label{eq:zero-net-force}
\int_{\Gamma_0}  \bld{f}({\bld x}) \mathrm{d}\bld{x} = 0.
\end{equation} 
We solved equations~(\ref{eq:int-one-lambda}) and (\ref{eq:zero-net-force}) numerically in Matlab by splitting the integration path into $N$ straight-line segments of equal length $\Delta s$ and replacing the line integrals in equations~(\ref{eq:int-one-lambda}) and (\ref{eq:zero-net-force}) by summation over the segments,        
\begin{equation}
\label{eq:discrete-equation}
\bld{u}_S(\bld{x}_i)+ \bld{U} = \frac{1}{4\pi\eta_m} \sum_{j = 1}^N \sum_{m=-M}^{M}  \balpha(\bld{x}_i - \bld{x}_{mj})\bcdot \bld{f}(\bld{x}_{j}) \Delta s,
\end{equation}
\begin{equation}
\label{eq:zero-net-force-discrete}
\sum_{j = 1}^N  \bld{f}(\bld{x}_{j}) = 0.
\end{equation}
In equations~(\ref{eq:discrete-equation}) and (\ref{eq:zero-net-force-discrete}) $\bld{x}_{i(j)}$ are the coordinates of the segments' midpoints.
In equation~(\ref{eq:discrete-equation}) we introduced the truncation parameter $M$ for the infinite sum over the wavelengths.

For a starting value of parameter $M$ (usually $M =10$), we ran computations for five different values of parameter $N$ in the range $300$--$1000$ and then extrapolated our results for the swimming velocity to $\Delta s \to 0$ ($N\to\infty$). We then gradually increased the value of $M$ and repeated the computations until the solution for $\bld{U}$ converged, showing changes smaller than $0.5\%$ with further increase of the number of terms in the sum over $m$. The computations required increasingly more terms in the sum over $m$ for large amplitudes ($bq > 1$) and large Saffman lengths ($\lambda/\ell_S \ll 1$) due to strong long-range hydrodynamic interactions between segments of the flagellum in this (nearly 2D) regime, and correspondingly large contribution to the flow field by the forces $\bld{f}(\bld{x}_{j})$ separated by multiple wavelengths along the flagellum.

We paid 
special attention to the diagonal term with  $m=0$ and $i=j$ in equation~(\ref{eq:discrete-equation}). This term gives the fluid velocity in the close proximity of a localized force 
$\bld{f}(\bld{x})$. The response tensor $\balpha(\bld{x})$  diverges due to logarithmic singularities in the functions $\alpha_\|(\bld{x})$ and $\alpha_\perp(\bld{x})$ in the limit $\bld{x}\to 0$ [see equations~(\ref{eq:alpha-parallel}) and (\ref{eq:alpha-perp})]. 
 In the close proximity of a localized force the fluid velocity is parallel to the force, and is, therefore, determined by the parallel component of the response function $\alpha_\|(\bld{x})$. We expanded $\alpha_\|(\bld{x})$ about $\bld{x}=0$ and performed integration analytically over $\Delta s$ in the vicinity of $\bld{x}=0$. Therefore, for the diagonal term on the RHS of equation~(\ref{eq:discrete-equation}), 
 which we denote as $\bld{{\cal A}}^{m=0}_{i=j}$, we have
\begin{eqnarray}
\bld{{\cal A}}^{m=0}_{i=j} (\bld{x}_i) &=& \frac{1}{4\pi\eta_m} \bld{f}(\bld{x}_{i})  \int_{-\Delta s/2}^{\Delta s/2} \alpha_\|( z ) \mathrm{d}z  \nonumber \\
 &=& \frac{1}{4\pi\eta_m} \bld{f}(\bld{x}_{i}) \, 2 \lim_{\varepsilon \to 0} \int_{\varepsilon}^{\Delta s /2}
\left [ \frac{1}{2}  - \gamma + \frac{2z}{3\ell_S} + \log \frac{2\ell_S}{z} \right ] \mathrm{d}z  \nonumber \\
&=& \frac{1}{4\pi\eta_m} \bld{f}(\bld{x}_{i}) \,  \Delta s \left [  \frac{3}{2} + \frac{\Delta s}{6\ell_S} - \gamma + \log\left(\frac{4\ell_S}{\Delta s} \right) \right ], 
\label{eq:diag-term-x}
\end{eqnarray}
where $\gamma = 0.5772$ is the Euler constant. 
\begin{figure}
\centerline{\includegraphics[width=14cm]{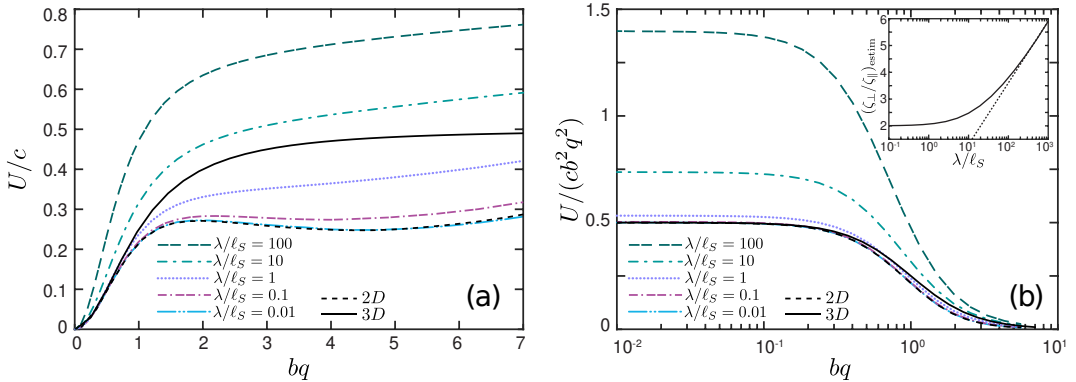}}
\caption{Calculated swimming speed vs. $bq$ for various ratios  $\lambda/\ell_S=2\pi/(q\ell_S)$ of an infinitely long {\it inextensible} flagellum (a) scaled by the wave speed $c$, and (b) scaled by $c (bq)^2$.
The colored (grey) curves are the results of our BEM computations described in the text.  The black dashed curve corresponds to the swimming speed in a purely 2D fluid, and the solid black curve is the local drag theory result of Gray and Hancock \citep{GrayHancock1955} for a flagellum in a 3D unbounded fluid oscillating with moderately large amplitude. The solid curve in the
inset in panel (b) represents an estimate for the ratio of the drag constants as a function of the scaled wavelength, as discussed in the text. The wave amplitude in the inset was set to $bq = 10^{-3}$, and the dotted line  corresponds to $y = \ln x + \mbox{const}$ 
for reference.}
\label{fig:U-inf-flagellum}
\end{figure}

Our computations confirm that an infinitely long flagellum has a non-vanishing component of the swimming velocity only along the $x$-axis, as expected by symmetry.  In figure~\ref{fig:U-inf-flagellum} we plot the swimming speed as a function of the dimensionless amplitude $bq$ for a range of wavelengths scaled by the Saffman length. In the limit of  
pure 2D hydrodynamics, 
which corresponds to large Saffman lengths (and small scaled wavelengths, $\lambda/\ell_S \ll 1$), the energy dissipation occurs primarily in the membrane, and the membrane's viscous drag on the flagellum makes the main contribution to the flagellum's propulsion. In this limit, our computations are in good agreement with the 2D problem of the Taylor swimming sheet, as expected because a Taylor `string' in a thin very viscous membrane is equivalent to the Taylor sheet in 3D bulk fluid.
For small amplitudes ($bq\ll 1$), we recover Taylor's leading order perturbative solution $U/c = (1/2) (bq)^2$. For larger amplitudes $bq$ (and $\lambda/\ell_S \ll 1$), our calculated swimming speed is in agreement with recent analytic and computational results of Sauzade and coworkers \citep{Sauzade2011}.

 In figure~\ref{fig:U-inf-flagellum} the black dashed curve corresponds to a flagellum swimming in a pure 2D membrane (no bulk fluid surrounding the membrane). We briefly outline our computations for this limiting case in Appendix~\ref{appA}. The computations are similar to the boundary integral approach demonstrated in \citep{Sauzade2011} with the only difference that we neglected the double layer potential contribution. 
The solid  black curve in figure~\ref{fig:U-inf-flagellum} corresponds to the local drag theory for an infinitely long flagellum passing waves of moderately large amplitudes in a 3D unbounded fluid \citep{GrayHancock1955}.   

As can be seen in figure~\ref{fig:U-inf-flagellum}, our BEM computations predict that  for wavelengths larger than the Saffman length ($\lambda/\ell_S > 1$) the swimming speed in a quasi-2D membrane exceeds that in purely 2D and 3D fluids. For qualitative explanation of this result we compare our BEM computations with the local drag model of  \citet{GrayHancock1955}. In the local drag approximation, one assumes that the viscous drag force on a small segment of the flagellum is proportional to the segment's velocity, and the total drag on the swimmer is a sum over these local drag forces. Thus, the local drag approximation does not explicitly take into account the long-range hydrodynamic interactions between distant segments of the flagellum. The local drag forces for the motion of a rod-like segment parallel and perpendicular to its geometrical axis are given by $F_\| = \zeta_\| v_\|$ and $F_\perp = \zeta_\perp v_\perp$, respectively, with the drag coefficients $\zeta_\|$ and $\zeta_\perp$. 

We expect our BEM results to be in qualitative agreement with the local drag approximation in the limit of $bq\ll1$ and $\lambda/\ell_S \gg 1$. For small amplitudes $bq$ the segments of an inextensible flagellum separated by large {\it contour} distances ($>\lambda$) do not come too close to each other, and for the wavelengths larger than the Saffman length the spatial decay of the flow field is faster ($\sim 1/r$), in comparison with slower (logarithmic) decay rate for $\lambda/\ell_S \ll 1$.  Thus in the regime of $bq\ll1$ and $\lambda/\ell_S \gg 1$, the cooperativity effect between distant segments of the flagellum is expected to be small. \citet{GrayHancock1955} obtained the swimming velocity of an infinitely long and thin flagellum to the leading order of amplitude $bq$, 
\begin{equation}
\label{eq:Gray-Hancock-velocity}
\frac{U}{c} = \frac{(bq)^2}{2} \left(\frac{\zeta_\perp}{\zeta_\|} - 1\right).
\end{equation}
In 3D, the ratio of the drag coefficients for an infinitely thin rod is $\zeta_\perp/\zeta_\| = 2$. For inclusions in quasi-2D membranes, the ratio $\zeta_\perp/\zeta_\|$ depends on the Saffman length. 
In the inset of figure~\ref{fig:U-inf-flagellum} we plot our BEM results for $(\zeta_\perp/\zeta_\|)_{\rm estim} \equiv 1 +  2U/(cb^2q^2) $ as a function of $\lambda/\ell_S$. According to  equation~\ref{eq:Gray-Hancock-velocity}, $(\zeta_\perp/\zeta_\|)_{\rm estim}$ should give us an estimate for the local drag anisotropy. For the plot in the inset we chose a small amplitude $bq = 10^{-3}$, when the comparison with the local drag calculation of Gray and Hancock is justified.  As can be seen in the inset of figure~\ref{fig:U-inf-flagellum}, the effective ratio  $(\zeta_\perp/\zeta_\|)_{\rm estim}$ grows logarithmically with $\lambda/\ell_S$ for $\lambda/\ell_S \gg 1$. 

This result is in qualitative agreement with the work of Levine and collaborators \citep{Levine_rodmobility2004}, some of which we summarize here.
Levine {\it et al.} studied the drag coefficients for a rod-like inclusion of length $L$ moving in a quasi-2D membrane, and
showed that for rod-like inclusions of lengths smaller than the Saffman length ($L/\ell_S \ll 1$), where the viscous dissipation occurs primarily in the membrane, the dependence of the drag coefficients on the size and orientation of the rod is weak: $\zeta_\perp/\zeta_\| \to 1$.  For longer rods with $L/\ell_S \gg 1$, the dissipation is governed by the 3D fluid surrounding the membrane, and the drag coefficients show a stronger dependence on the size of the rods.  Levine {\it et al.} found that the drag coefficient $\zeta_\|$ for a thin rod in a quasi-2D membrane is qualitatively similar to that in 3D and is given by
\begin{equation}
\zeta_\| = \frac{2\pi \eta L}{\ln\left(0.43 L/\ell_S\right)}.
\end{equation}
However, the dependence of $\zeta_\perp$ on $L$ in a quasi-2D membrane is very different from that in 3D: 
\begin{equation}
\zeta_\perp=2\pi\eta L;
\end{equation}
$\zeta_\perp$ depends on $L$ linearly, 
without the logarithmic factor in the denominator.
The linear dependence of $\zeta_\perp$ on $L$ indicates 
the local character of the drag and the effective absence of 
hydrodynamic interactions between different sections of the rod. 

As emphasized by \citet{Levine_rodmobility2004}, this behavior of $\zeta_\perp$ 
arises from the incompressibility 
of the membrane, $\bnabla_\perp\bcdot\boldsymbol{v}_\perp = 0$, where the symbol $\perp$ denotes differentiation  `in-plane' and the components of the velocity field in the plane. In the case of a  rod moving perpendicular to its long axis in a 3D fluid,  the fluid can flow past the rod by moving over and under it. In this flow pattern, the in-plane part of the incompressibility condition does not vanish: $\partial_x v_x + \partial_y v_y\neq0$. 
A 2D version of such a flow (its projection on the $xy$-plane) is impossible due to the membrane incompressibility.
In a quasi-2D membrane, the fluid moves the long way around the rod, and the flow extends over distances comparable to the largest rod dimension $L$. 
The membrane incompressibility constraint only affects the perpendicular drag on a filament. A segment of filament being dragged parallel to its long axis does not produce divergent flows in a simple fluid, and thus the flow character is unchanged by the presence of a membrane.
 
Therefore, for long wavelengths ($\lambda/\ell_S \gg1$), the membrane incompressibility is expected to lead to a logarithmic growth of the flagellum's effective drag anisotropy, $\zeta_\perp/\zeta_\| \propto \log(\lambda/\ell_S)$.  An organism that relies on the drag anisotropy for propulsion would achieve greater swimming speeds in a quasi-2D membrane than in pure 2D or 3D fluids, as is confirmed by our BEM computations. 

\section{Finite length flagellum in a quasi-2D membrane} 
\label{sec:finite}
\begin{figure}
\centerline{\includegraphics[width=7cm]{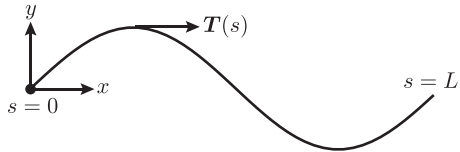} }
\caption{In the body frame the `head' of the flagellum is motionless and is placed at the origin of the coordinate system. The position of a material particle is determined by the arc length $s$ measured from the left end of the flagellum. The tangent vectors $\bld{T}(s)$ describe the instantaneous shape of the flagellum. The flagellum propagates planar sinusoidal waves to the right.}
\label{fig:sinusoidal}
\end{figure}

We also applied the BEM to the case of an inextensible, headless, infinitely thin flagellum of finite length. As in the case of an infinitely long flagellum, the motion of the swimmer is prescribed by a sinusoidal modulation, $y(s, t) = b\sin(q x(s) - \omega t + \phi_0)$. Here $s$ is the arc length along the flagellum measured from the flagellum's hypothetical `head' and $\phi_0$ is the initial phase. At every time instant, the shape of the flagellum is described by the curve $\bld{X}(s, t)$,
where $\bld{X}(s, t) = (X(s), Y(s)) = (x(s), y(s, t) - y(0, t))$ (see figure~\ref{fig:sinusoidal}). The unit tangent to the curve is $\bld{T}(s)  = (\mathrm{d} X/\mathrm{d} s, \mathrm{d} Y/\mathrm{d} s)$. 

In the frame of the flagellum, a material point at position $s$ moves with velocity $\bld{u}_S(s, t) = \partial \bld{X}(s, t)/\partial t$. As was demonstrated by \citet{Higdon1979},  in the case of transverse waves propagating along the flagellum, the material particle's velocity can be calculated in a different manner.  In a reference frame moving with the wave, the shape of the flagellum is given by $\bld{X}^w(s - C t)$, where $C$ is the arcwise speed that we introduced in equation~(\ref{eq:Q-speed}).  Following Higdon, we note that 
 \begin{equation}
 X^w(s + \Lambda) = X^w(s) + \lambda, \qquad Y^w(s + \Lambda) = Y^w(s), 
 \end{equation}
 where $\Lambda$ is the arcwise wavelength (see equation~(\ref{eq:arc-Lambda})) and $\lambda$ is the linear wavelength. 
 The tangential vectors are identical in the body and wave frames since the wave frame simply translates with respect to the body frame and does not undergo rotation. Thus, $\bld{T}^w(s - C t) = \bld{T}(s- C t)$. The velocity of a material particle at $s$ in the wave frame is calculated as $\bld{u}^w (s, t) = \partial \bld{X}^w(s - C t)/\partial t = - C  \partial \bld{X}^w/\partial  s = - Q  \bld{T}(s- C t)$. The velocity of the `head' in the wave frame is 
$ - C  \bld{T}(s - C t)\Big{|}_{s=0} = - C  \bld{T}( - C t)$. Therefore, the wave frame translates with respect to the body frame with velocity $C \bld{T}( - C t)$. Therefore, we can calculate the velocity of the material point $s$ in the body frame as 
\begin{equation}
\bld{u}_S(s, t) = C \bld{T}( - C t)  -  C  \bld{T}(s - C t).
\end{equation}
 The material velocity with respect to the fluid at infinity is then
 \begin{equation}
 \bld{u}_S(s, t)   + \bld{U}(t) + \bld{\Omega}(t) \times \bld{X}(s, t), 
 \end{equation}
 where $\bld{U}(t)$ and $\bld{\Omega}(t)$ are the translational and angular velocities of the flagellum. 

Similarly to our treatment of an infinitely long flagellum in \S\ref{Sec:infinite-flagellum}, we model the fluid velocity due to the flagellum's motion as a linear superposition of velocities due to a force density (see equation~(\ref{eq:membrane-velocity})). Now the path of integration $\Gamma$ stands for the curve $\bld{X}(s, t)$ describing the instantaneous shape of the flagellum. The instantaneous swimming and angular velocities and the force density are calculated from the coupled integral equations for a no-slip boundary condition on the surface of the flagellum and the requirements of zero net force and torque on the flagellum,
\begin{equation}
\bld{u}_S(s, t)  +   \bld{U}(t)  + \bld{\Omega}(t) \times \bld{X} =  \frac{1}{4\pi\eta_m} \int_\Gamma 
\balpha(\bld{X} -\bld{X}^\prime)\bcdot \bld{f}({\bld X^\prime}) \mathrm{d}{\bld X}^\prime,
\end{equation}
\begin{eqnarray}
 \int_\Gamma  \bld{f}({\bld X}) \mathrm{d}{\bld X} &= &0,\\
 \int_\Gamma  {\bld X} \times \bld{f}({\bld X}) \mathrm{d}{\bld X} &= &0,
\end{eqnarray}
where ${\bld X} \equiv {\bld X}(s, t)$ and ${\bld X}^\prime \equiv {\bld X}(s^\prime, t)$.

Similar to the approach discussed in \S\ref{Sec:infinite-flagellum}, we solved the discretized version of these equations for the instantaneous velocities $\bld{\Omega}(t)$, $\bld{U}(t)$ and the force density $\bld{f}({\bld X})$. We normally calculated the angular and swimming velocities for about 60--80 snapshots per one period of oscillation and averaged them over one cycle of motion, $\langle \Omega \rangle = \frac{1}{T} \int_0^T \Omega(t)  \mathrm{d}t \approx  \frac{1}{N_T} \sum_{i=1}^{N_T} \Omega(t_i)$ and 
$\langle \bld{U} \rangle = \frac{1}{T} \int_0^T \bld{R}(t)\cdot \bld{U}(t)  \mathrm{d}t \approx  \frac{1}{N_T} \sum_{i=1}^{N_T}\bld{R}(t_i)\cdot \bld{U}(t_i) $, where $\bld{R}(t)$ is the rotation operator that transforms the swimming velocity vector to the initial coordinate system $\bld{X}(s, 0)$, which is motionless with respect to the fluid at infinity, and $N_T$ is the number of snapshots per one period.

\begin{figure}
\centerline{ \includegraphics[width=11cm]{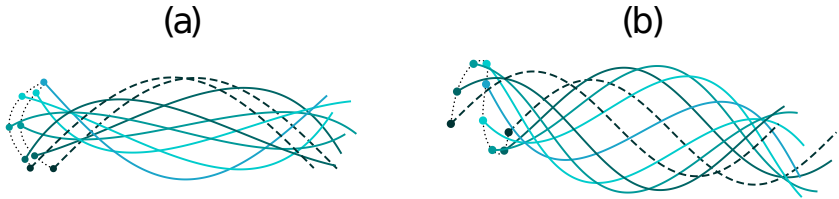}}
\caption{Propulsion of a flagellum during one period of oscillations. The time instants are separated by one-eighth of the period. Flagella drawn with dashed lines correspond to the time moments $t=0$ and $t=T$. The thin dotted line is the trajectory of the flagellum's `head' over one cycle of motion. The flagellum length is (a)  $N_\Lambda=0.5$,  (b) $N_\Lambda=1$.  In both (a) and (b) the amplitude and the wavelength were set to  $bq =1$ and $\lambda/\ell_S = 1$, respectively.}
\label{fig:snapshots}
\end{figure}

During one cycle of motion, a flagellum of finite size moves in a tortuous fashion that involves pitching (rotation of the swimmer's centerline with respect to the initial direction of wave propagation) and drifting (motion of the flagellum's center of mass perpendicular to the net swimming direction), see figure \ref{fig:snapshots}.

\begin{figure}
\centerline{\includegraphics[width=13cm]{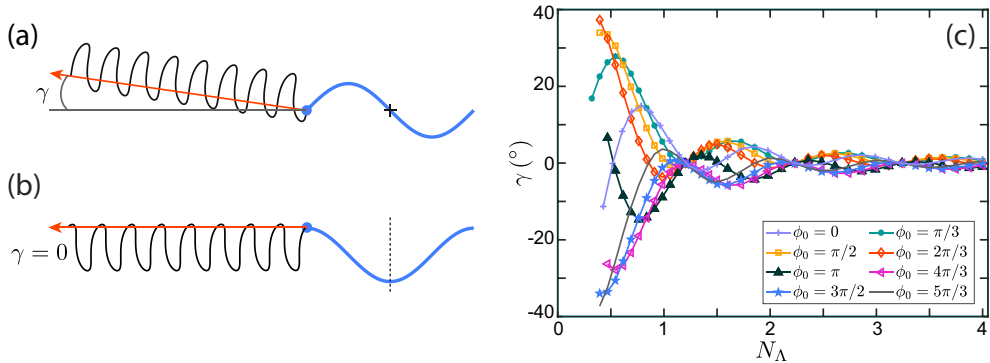}}
\caption{The swimming trajectory of the head (black) for a flagellum (a) starting its motion in an odd configuration, and (b)  starting its motion in an even configuration. The arrow shows the direction of the average swimming velocity. (c) The angle (in degrees) of the average swimming velocity with respect to $(-x)$-axis as a function of the flagellum's contour length $N_\Lambda$. We set $bq =1$ and $\lambda/\ell_S = 1$.}
\label{fig:direction_of_swimming}
\end{figure}

Our computations predict that $\langle \Omega \rangle =0$, and a flagellum of finite size swims in a straight line at an angle $\gamma$ with respect to $(-x)$-axis, where $\gamma = \arctan(\langle U_y \rangle/ |\langle U_x \rangle|)$. The magnitude and sign of $\gamma$ depend on the flagellum's contour length, the wave amplitude and the phase constant $\phi_0$. For particular values of $\phi_0$, a flagellum of a given contour length can assume an even or odd configuration at time $t=0$. In the even configuration (as illustrated in figure \ref{fig:direction_of_swimming}b) the flagellum has reflection symmetry with respect to the vertical dashed line that passes through the flagellum's center of mass. In the odd configuration the shape of the flagellum has point symmetry about the center of the flagellum (marked by cross hairs in figure \ref{fig:direction_of_swimming}a).

\citet{Koehler2012}, in work on the swimming of finite-length flagella in a Newtonian 3D fluid, used symmetry arguments to prove that a flagellum that starts its motion from an even configuration swims along its centerline in the direction opposite to initial wave propagation, and does not drift away from the $(-x)$-direction. \citet{Koehler2012} pointed out that the mirror reflection about the vertical line is equivalent to the time reversal, with the time-reversed swimming velocity $\bld{U}_{-t} = -\bld{U}_t$.  Therefore, the instantaneous swimming velocity in an even configuration must be identical to the mirror image of the time-reversed velocity. This condition requires the $y$- component of the swimming velocity be equal to zero. Thus, a flagellum starting its motion from an even configuration, has vanishing mean drift over one cycle of motion. A flagellum starting its motion in an odd configuration, will reach an even configuration after a quarter of a period. The reorientation angle acquired by the flagellum by $t=T/4$  determines the flagellum's swimming direction. The reorientation angle for a flagellum starting its motion in an odd configuration is the largest since it takes the longest amount of time for such a flagellum to reach an even configuration.
In figure~\ref{fig:direction_of_swimming} we plot $\gamma$ as a function of the flagellum's contour length $N_\Lambda \equiv L/\Lambda$ for various values of the phase constant $\phi_0$.
A similar swimming pattern was reported by \citet{Peng2016} in the work on flagella locomotion in granular media.  

\begin{figure}
\centerline{\includegraphics[width=13.5cm]{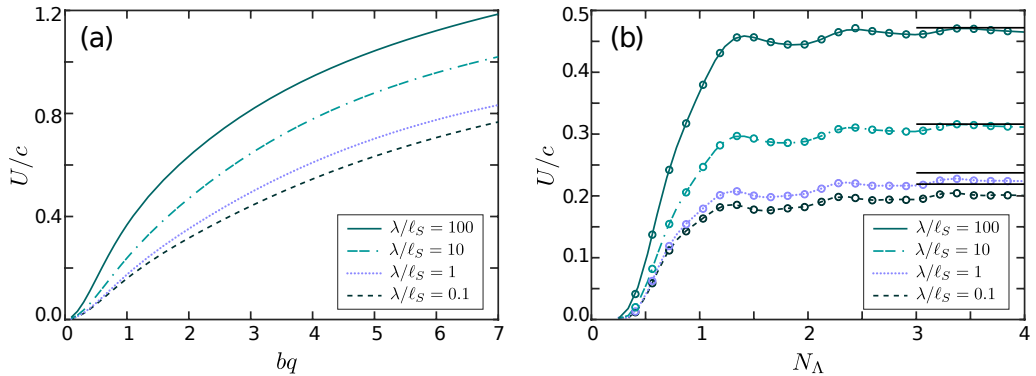}}
\caption{Calculated swimming speed scaled by the wave speed of an inextensible, headless finite-length flagellum (a) as a function of parameter $bq$ for a flagellum length equal to one arcwise wavelength, $N_\Lambda = 1$, (b)  as a function of the scaled flagellum length $N_\Lambda$ for $bq = 1$. In (b) the circles represent calculations based on the Lorentz reciprocal theorem (see equation~(\ref{eq:Lorentz-for-swimmer})). The horizontal black lines are asymptotes for the swimming velocities in the limit $N_\Lambda \to \infty$ calculated using the method described in \S\ref{Sec:infinite-flagellum}.}
\label{fig:swimming-U}
\end{figure}

In figure~\ref{fig:swimming-U}a we plot the swimming speed averaged over one period of oscillations, $U = |\langle \bld{U} \rangle|$, as a function of a dimensionless parameter $bq$ for the flagellum contour length equal to one arcwise wavelength, $N_\Lambda =1$.  Two competing mechanisms influence the swimming speed of the flagellum. On the one hand, larger values of $bq$ correspond to steeper angles between the flagellum and the direction of wave propagation and, therefore, a stronger propulsion force. On the other hand, for larger $bq$ values the segments of the flagellum come closer to each other. The hydrodynamic interactions between the segments tend to slow down the swimmer.  
The hydrodynamic interactions are stronger in the limiting case of 2D hydrodynamics (small $\lambda/\ell_S$ ratios) due to a slow, logarithmic spatial decay rate of the flow field. In the opposite limit of large $\lambda/\ell_S$ ratios, our calculations do not reproduce Higdon's results for the swimming speed in a purely 3D fluid \citep{Higdon1979}. Being qualitatively similar to Higdon's prediction, our calculations show much larger swimming speeds for $\lambda/\ell_S \gg 1$. As we discussed at the end of \S\ref{Sec:infinite-flagellum}, the incompressibility of the membrane sets a constraint on the fluid dynamics that leads to an effective drag anisotropy that grows logarithmically as a function of $\lambda/\ell_S$ for $\lambda/\ell_S \gg 1$.  The enhanced drag anisotropy in a quasi-2D membrane is responsible for larger swimming speeds in quasi-2D membranes (in comparison with  pure 2D or 3D fluids). 

In figure~\ref{fig:swimming-U}b we plot the swimming speed as a function of the scaled flagellum length $N_\Lambda$ for $bq=1$.  For $N_\Lambda  < 1$ the flagellum performs large yawing motion that is inefficient for swimming (see figure~\ref{fig:snapshots}a). For larger values of $N_\Lambda $ the long-range hydrodynamic interactions taper off the growth of the swimming speed, and the speed approaches the values found for an infinitely long flagellum (shown as dotted black horizontal lines in figure~\ref{fig:swimming-U}b). 
The `bumps' in the curves reflect smaller yawing of the flagellum for some values of $N_\Lambda$. 

To find the flagellum motion that is optimal in terms of the power consumption, we calculated the swimming efficiency. As discussed in \citep{Koehler2012}, there are multiple efficiency metrics.  Here we calculated the efficiency as the ratio of the power required to pull the flagellum through the fluid at its average swimming speed, $F_{\rm pull}  U =  (1/ \mu_\| )  U^2$, to the average power $\langle P \rangle$ consumed by the swimmer over one period of motion, 
\begin{equation}
\eta = \frac{(1/ \mu_\| ) U^2}{\langle P \rangle}. 
\end{equation}
Here $ \mu_\| $ is the flagellum mobility for the translational motion along the direction of wave propagation, averaged over one period.  The average power consumption is 
\begin{equation}
{\langle P \rangle} =  \frac{1}{T} \int_0^T \mathrm{d} t \int_\Gamma  \mathrm{d}{\bld X} \bld{f}({\bld X}, t) \cdot \bld{u}_S({\bld X}, t). 
\end{equation}
We calculated the mobility and the power consumption numerically using the BEM. 

\begin{figure}
\centerline{\includegraphics[width=14cm]{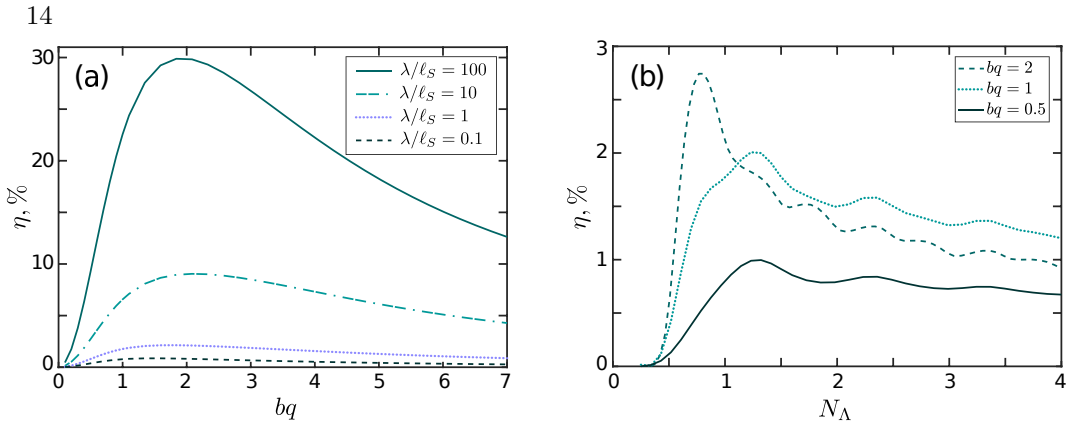}}
\caption{Calculated flagellum efficiency of an inextensible finite-length flagellum (a) as a function of the flagellum amplitude $bq$  for $N_\Lambda = 1$, (b) as a function of $N_\Lambda$  for $\lambda/\ell_S = 1$.}
\label{fig:efficiency}
\end{figure}

Our calculated flagellum efficiency is in qualitative agreement with \citet{Higdon1979}.  In figure~\ref{fig:efficiency}a we plot the efficiency as a function of the amplitude $bq$ for a few values of the scaled wavelength $\lambda/\ell_S$ for a flagellum of length $N_\Lambda = 1$. For smaller values of $bq$ the segments of the flagellum have small angles with respect to the direction of wave propagation and produce a weak thrust.  For larger values of $bq$ the flagellum `shrinks' along the $x$-axis, and the stronger interference between the segments of the flagellum leads to a decrease in efficiency. The efficiency increases with $\lambda/\ell_S$ due to the reduced role of long-range hydrodynamics on length scales exceeding the Saffman length $\ell_S$. The maximum efficiency falls at  $bq =$ 1.4 -- 1.8. 

In figure~\ref{fig:efficiency}b we plot the efficiency as a function of the flagellum length $N_\Lambda$ for a wavelength $\lambda/\ell_S = 1$. For small values of $N_\Lambda$ the swimming of the flagellum is inefficient due to an excessive yawing motion and a weak overall thrust (see figure~\ref{fig:snapshots}a). The efficiency reaches a maximum at  $N_\Lambda$ = 0.8 -- 1.4  and then decreases with further growth of $N_\Lambda$ due to interference between the crests of the flagellum. The interference is stronger for larger amplitudes $bq$ since the crests are closer to each other, and the efficiency drops off more abruptly from its optimal value for larger values of $bq$. The secondary maxima correspond to a smaller yawing and larger propulsion for various values of $N_\Lambda$.  Figure~\ref{fig:snapshots} demonstrates that the flagellum travels a considerable distance along the $y$-axis while making moderate overall progress along the $x$-axis. 

In the following section we discuss an alternative computational approach to finding the swimming velocities using the Lorentz reciprocal theorem.
\section{Lorentz reciprocal theorem for a quasi-2D membrane}
\label{Sec:Lorentz}
Finding the analytical solution for the swimming velocity can be a daunting task. One of the major difficulties is that one needs to solve Stokes equations with time dependent no-slip boundary conditions on the surface of the swimmer. 
\citet{StoneSamuel1996} offered an elegant way to find the swimming velocity using the Lorentz reciprocal theorem (LRT) \citep{happel1965low}. For a fluid in 3D the Lorentz reciprocal theorem states that if there are two solutions to Stokes equations and the incompressibility condition with the velocity fields and the stress tensors $(\boldsymbol{v},\ \boldsymbol{\sigma})$ and $(\boldsymbol{v}^\prime,\ \boldsymbol{\sigma}^\prime)$, respectively, that satisfy the same boundary conditions at infinity, than for a volume of fluid $V$ bounded by surface $S$, we have
\begin{equation}
\oint_S \boldsymbol{v}\cdot \boldsymbol{\sigma}^\prime \cdot \boldsymbol{n}\, \mathrm{d} S= \oint_S \boldsymbol{v}^\prime\cdot \boldsymbol{\sigma} \cdot \boldsymbol{n} \, \mathrm{d} S,
\end{equation}
where $\boldsymbol{n}$ is the outward normal to the surface $S$. 

Here we formulate the LRT approach for a finite swimmer confined to a quasi-2D membrane. Let $\boldsymbol{v}$ and $\boldsymbol{\sigma}$  be the membrane velocity and the stress fields for the swimming problem.  These velocity and stress fields  are solutions of the Stokes equations and the incompressibility condition, equation~(\ref{eq:Stokes1}). They also satisfy the conditions of zero net force and torque on the swimming body, equations~(\ref{eq:net-force1}) and (\ref{eq:net-torque}). For the reciprocal solution of equation~(\ref{eq:Stokes1}) we choose the membrane velocity ${\boldsymbol{v}}^\prime$ and stress ${\boldsymbol{\sigma}}^\prime$  fields due to an inactive object of the same shape as the swimmer and being dragged as a solid body with constant translational velocity ${\boldsymbol{U}}^\prime$.   
\begin{figure}
\centerline{\includegraphics[width=7cm]{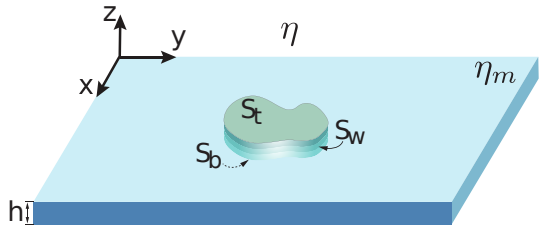}}
\caption{A region in a quasi-2D membrane of the same geometry as a swimmer. The integration surface $S$ in equation~(\ref{eq:net-force}) is comprised of the curvy wall $S_{w}$ embedded in the membrane, the top ($S_t$) and the bottom ($S_b$) surfaces of the membrane that are in contact with the bulk fluid.}
\label{fig:inclusion-in-membrane}
\end{figure}

When the condition $\bnabla \cdot \boldsymbol{\sigma} = 0$ is relaxed (see equation~(\ref{eq:Stokes2})), a more general form of the Lorentz reciprocal theorem is \citep{kimbook}
\begin{eqnarray}
\label{eq:Lorentz}
\oint_S {\boldsymbol{v}}^\prime \cdot (\boldsymbol{\sigma} \cdot \boldsymbol{n}) \mathrm{d} S  - \int_V {\boldsymbol{v}}^\prime \cdot (\bnabla \cdot \boldsymbol{\sigma}) \mathrm{d} V =
\oint_S \boldsymbol{v} \cdot ({\boldsymbol{\sigma}}^\prime \cdot \boldsymbol{n}) \mathrm{d} S - \int_V \boldsymbol{v} \cdot (\bnabla \cdot {\boldsymbol{\sigma}}^\prime) \mathrm{d} V,
\end{eqnarray}
where $V$ is the swimmer's volume bounded by the surface $S$. Here we treat the volume occupied by the swimmer as being equivalent to the fluid domain of the same shape as the swimmer's and having the same velocity distribution as that of the material particles of the swimmer. 

Let us consider the first term on the LHS of equation~(\ref{eq:Lorentz}):
\begin{eqnarray}
\oint_S {\boldsymbol{v}}^\prime \cdot (\boldsymbol{\sigma} \cdot \boldsymbol{n}) \mathrm{d} S = {\boldsymbol{U}}^\prime \cdot 
\int_{S_{w}} \mathrm{d} \boldsymbol{F},
\end{eqnarray}
where we took into account that ${\boldsymbol{v}}^\prime={\boldsymbol{U}}^\prime$ is a constant vector at the surface of the domain (uniform translation). Also, since $\boldsymbol{\sigma}$ does not have $z$- components, only $ \boldsymbol{\sigma}\cdot \boldsymbol{n}  \mathrm{d} S =  \mathrm{d} \boldsymbol{F}$ on the curvy wall of the domain, $S_w$ (see figure~\ref{fig:inclusion-in-membrane}), will make a non-zero contribution. 

Taking into account equation~(\ref{eq:Stokes2}), the second term on the LHS of equation~(\ref{eq:Lorentz}) can be rearranged as
\begin{eqnarray}
 -\int_V {\boldsymbol{v}}^\prime  \cdot (\bnabla \cdot \boldsymbol{\sigma} ) \mathrm{d} V = {\boldsymbol{U}}^\prime  \cdot \int_V \frac{2\boldsymbol{f}}{h} \mathrm{d} V 
 = {\boldsymbol{U}}^\prime \cdot \int_{S_{t, b}}  2\boldsymbol{f} \, \mathrm{d} S,
\end{eqnarray}
where we take into account that the traction forces $2\boldsymbol{f}$ due to the fluid flows in the surrounding fluid act on the flat sides of the domain $S_{t,b}$ (see equation~(\ref{eq:Stokes2}) and  figure~\ref{fig:inclusion-in-membrane}), and $dV = h dS$.
Therefore, the LHS of equation~(\ref{eq:Lorentz}) becomes:
\begin{eqnarray}
{\boldsymbol{U}}^\prime \cdot \left(\int_{S_{w}} \mathrm{d} \boldsymbol{F} +  \int_{S_{t,b}} 2\boldsymbol{f} \, \mathrm{d} S \right)= {\boldsymbol{U}}^\prime \cdot {\boldsymbol F} = 0, 
\label{eq:net-force}
\end{eqnarray}
where ${\boldsymbol F}$ is the net force on the swimmer and is equal to zero.

Similarly, for the terms on the RHS of equation~(\ref{eq:Lorentz}) we have
\begin{eqnarray}
\oint_S \boldsymbol{v} \cdot ({\boldsymbol{\sigma}}^\prime \cdot \boldsymbol{n}) \mathrm{d} S 
- \int_V \boldsymbol{v} \cdot (\bnabla \cdot {\boldsymbol{\sigma}}^\prime ) \mathrm{d} V &=&
 \int_{S_{w}}\boldsymbol{v}\cdot ({\boldsymbol{\sigma}}^\prime \cdot \boldsymbol{n}) \mathrm{d} S  + \int_{S_{t, b}} \boldsymbol{v} \cdot (2 {\boldsymbol{f}}^\prime) \, \mathrm{d} S \nonumber \\
 &=& \oint_S \boldsymbol{v} \cdot \mathrm{d} {\boldsymbol F}^\prime,
 \label{eq:RHS-of-Lorentz}
\end{eqnarray}
where in the last line of equation~(\ref{eq:RHS-of-Lorentz}) we merged two terms into one integral over the total surface of the swimmer, and $\mathrm{d} {\boldsymbol F}^\prime$ denotes an elementary traction force on the inactive `swimmer' being dragged with constant velocity. Thus, the Lorentz reciprocal relation, equation~(\ref{eq:Lorentz}), assumes a compact form,
\begin{equation}
\label{eq:Lorentz-compact}
0 =  \oint_S \boldsymbol{v} \cdot \mathrm{d} {\boldsymbol F}^\prime.
\end{equation}
Decomposing the surface velocity of the swimmer into the translational $\boldsymbol{U}(t)$ and the surface disturbance  $\boldsymbol{u}_S(t)$ velocities, $\boldsymbol{v}(t) = \boldsymbol{U}(t)+ \boldsymbol{u}_S(t)$,
we rewrite the Lorentz reciprocal relation in the form
\begin{equation}
\label{eq:Lorentz-for-swimmer0}
{\boldsymbol{F}}^\prime(t) \cdot \boldsymbol{U}(t) = - \oint_{S(t)} \boldsymbol{u}_S \cdot \mathrm{d} {\boldsymbol F}^\prime,
\end{equation}
similar to the equation derived by \citet{StoneSamuel1996} for a swimmer in a 3D fluid.  In equation~(\ref{eq:Lorentz-for-swimmer0}) the integration is performed over the instantaneous surface area of the swimmer. 
The generalization of (\ref{eq:Lorentz-for-swimmer0}) for the motion that involves rotation is
\begin{equation}
\label{eq:Lorentz-for-swimmer}
{\boldsymbol{F}}^\prime(t) \cdot \boldsymbol{U}(t) + {\boldsymbol{L}}^\prime (t)  \cdot \boldsymbol{\Omega}(t)= - \oint_{S(t)} \boldsymbol{u}_S \cdot \mathrm{d} {\boldsymbol F}^\prime,
\end{equation}
where ${\boldsymbol{L}}^\prime(t)$ is the torque applied to the inactive inclusion and $\boldsymbol{\Omega}(t)$ is the swimmer's angular velocity.

The Lorentz reciprocal relation, equation~(\ref{eq:Lorentz-for-swimmer}), is particularly useful for computation of the swimmer's translational and rotational velocities, $\boldsymbol{U}(t)$  and 
$\boldsymbol{\Omega}(t)$, when the stress tensor of the reciprocal problem (motion of an inactive body) is known. Unfortunately, it is also difficult to solve the reciprocal problem analytically for an inclusion of an arbitrary shape in a quasi-2D membrane, since the coupling with the bulk fluid makes the problem essentially three-dimensional.  When the reciprocal solution is not available, equation~(\ref{eq:Lorentz-for-swimmer}) can serve as an alternative computational path for finding the swimming velocity. 

As a test of equation~(\ref{eq:Lorentz-for-swimmer}), we found the swimming velocities of a finite inextensible flagellum as described in \S\ref{sec:finite}.
We solved the reciprocal problem numerically by finding the force densities $\bld{f}({\bld x}, t)$  for the uniform rotation of the flagellum about the $z$-axis and for the translational motion of the flagellum along the $x$- and $y$- axes for multiple  flagellum conformations corresponding to various time instants of the swimming cycle. We then solved the resulting system of three equations (\ref{eq:Lorentz-for-swimmer}) for the instantaneous angular velocity and $x$- and $y$- components of the translational velocity and found the average swimmer's speed over one period of oscillations. In figure~\ref{fig:swimming-U}b circles superimposed on the curves show the swimming velocities obtained using the Lorentz equation~(\ref{eq:Lorentz-for-swimmer}). 

\subsection{The two-dimensional squirmer example}
The Lorentz reciprocal theorem can significantly simplify computations in the case of tangential deformations of the swimmer's body, when the overall shape of the critter remains unchanged. In this case the reciprocal problem can be solved for a single time instant, and the instantaneous swimming velocity can then be found from equation~(\ref{eq:Lorentz-for-swimmer}) by plugging in the time-dependent surface disturbance velocity $ \bld{u}_S(t)$.
\begin{figure}
\centerline{\includegraphics[width=4cm]{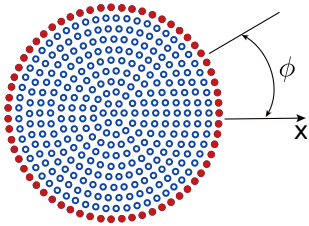}}
\caption{The flow field due to a squirmer is modeled as a superposition of the flow fields due to point-like forces (blobs). The blobs on the circumference (red) move with tangential velocity  $u_S(\phi) = B_1\sin\phi + B_2 \sin (2 \phi)$ in the body frame of the squirmer. The blobs in the interior of the squirmer (blue) are motionless in the body frame.}
\label{fig:discretized_disk}
\end{figure}

As an example, we consider a two-dimensional version of a squirmer, a critter that propels itself by beating its multiple hair-like appendages (cilia) in a periodic fashion. The periodic motion of the cilia carpet can be modeled by prescribing a velocity field on the surface of the squirmer. In a minimal model of a 2D squirmer, a disk-like body of radius $a$ is propelled due to a tangential disturbance velocity $u_S(\phi) = B_1\sin\phi + B_2 \sin (2 \phi)$ on the disk circumference (see figure~\ref{fig:discretized_disk}), where $B_1$ and $B_2$ are constants \citep{Blake1971, Papavassiliou2015}. The material points in the interior of the disk are motionless in the frame of the squirmer.  When the constants $B_1$ and $B_2$ have the same sign, they describe a contractile swimmer. Otherwise, the model corresponds to an extensile swimmer. 

In the limit of $a/\ell_S \ll 1$ the reciprocal problem for a disk was solved by \citet{Saffman1976} in the studies related to the Stokes paradox and a particle mobility in a quasi-2D fluid. In Appendix~\ref{appB} we outline calculations for the swimming velocity of the squirmer in the limit $a/\ell_S \ll 1$ using Saffman's solution for the reciprocal stress tensor $\boldsymbol{\sigma}^\prime$. The LRT reproduces the known swimming speed $U = B_1/2$, for a 2D squirmer in the limiting case of pure 2D hydrodynamics. 

Since the analytical solution for the reciprocal problem for a disk of arbitrary radius $a/\ell_S$ is not readily available, we found the reciprocal stress $\boldsymbol{\sigma}^\prime$ numerically by adopting the method of regularized Stokeslets (RS) for a quasi-2D membrane developed by \citet{Camley2013} in their work on mobility of inclusions in a quasi-2D membrane.  In \citep{Camley2013} the flow field due to a moving inclusion is modeled as a superposition of the flow fields due to point-like forces (blobs) tiling the disk area (see figure~\ref{fig:discretized_disk}),
\begin{equation}
\label{eq:fluidV}
u_\alpha ({\bld x}) = \sum_{i=1}^N \alpha_{\alpha\beta}({\bld x} - {\bld x}_i) f^\prime_\beta({\bld x}_i),
\end{equation}
where $\alpha, \beta = x, y$; $N$ is the total number of blobs; ${\bld x}_i$  is the in-plane coordinate of the $i$-th blob; $\alpha_{\alpha\beta}$ is the Levine-MacKintosh response function (see equations~\ref{eq:response-functions}); and $f^\prime_\beta({\bld x}_i)$ is the unknown force distribution. 

In the reciprocal problem the disk is being pulled through the membrane as a solid object with some given velocity ${\bld U}^\prime$. The force distribution $f^\prime_\beta({\bld x}_i)$ over the blobs is found by imposing a no-slip boundary condition on each 
blob,
\begin{equation}
U^\prime_\alpha= \sum_{j=1}^N \alpha_{\alpha\beta}({\bld x}_i - {\bld x}_j) f^\prime_\beta({\bld x}_j).
\end{equation}
Due to the squirmer's reflection symmetry about the $x$-axis, the rotational motion of the squirmer in an unbounded domain is ruled out, and the swimming velocity ${\bld U}$ can be found from the discretized version of equation~(\ref{eq:Lorentz-for-swimmer0}),
\begin{equation}
\label{eq:discrete-squirmer}
\left( \sum_{i=1}^N {\bld f}^\prime ({\bld x}_i) \right) \cdot {\bld U}  = - \sum_{\rm blobs\ on\ rim} \bld{u}_S(\bld{x}_j) \cdot {\bld f}^\prime ({\bld x}_j).
\end{equation}
The summation on the RHS of equation~(\ref{eq:discrete-squirmer}) is carried out only over the blobs on the squirmer's circumference since the inner blobs are motionless in the critter's frame.
\begin{figure}
\centerline{\includegraphics[width=13cm]{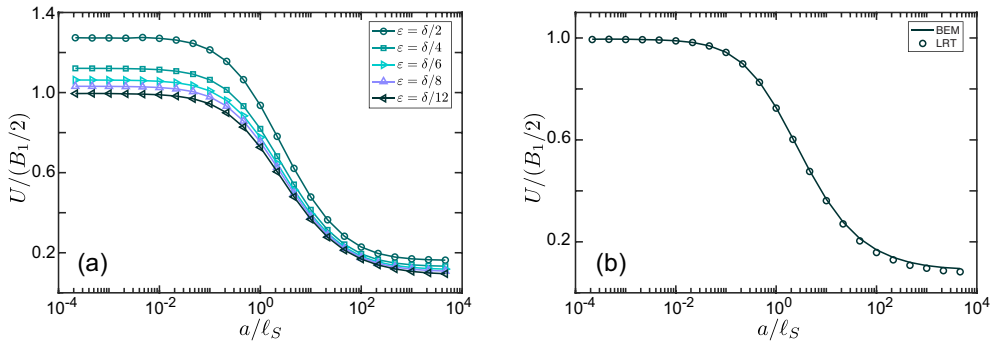}}
\caption{ 
Lorentz reciprocal theorem results for the squirmer swimming speed scaled by $B_1/2$ as a function of scaled squirmer radius $a/\ell_S$ (a) for various values of the regularization parameter $\varepsilon$, where $\delta$ is the distance between the centers of neighboring blobs,  (b) for $\varepsilon = \delta/12$ (circles).  The black curve corresponds to the BEM calculations for $\varepsilon = \delta/12$. The swimming speed is independent of the ratio $B_2/B_1$.}
\label{fig:epsilon_dependence}
\end{figure}
In the RS method the logarithmic singularity of the membrane response functions for $\kappa\to 0$ is eliminated by the regularization (smoothing) process that involves integration of the response function over the blob envelope function centered at $\kappa=0$. 
\citet{Camley2013} selected a Gaussian function for the regularization. The width of the Gaussian is controlled by an auxiliary parameter $\varepsilon$. 
Camley and Brown set $\varepsilon = \delta/2$, where $\delta$ is the distance between the centers of adjacent blobs, calculated the inclusion mobilities for several values of $\delta$ in the range $(0.03$--$0.07)a$, and extrapolated the results to the limit $\delta\to 0$. While the numerical calculations for the inclusion mobilities are only weakly dependent on the choice of $\varepsilon$ in our case of a swimming squirmer, the swimming velocities are more sensitive to the choice of the regularization parameter $\varepsilon$, since
 it effectively determines the thickness of the squirmer's deforming outer ring and its permeability, and therefore becomes a physical parameter.  

In figure \ref{fig:epsilon_dependence}a we plot the LRT results for the scaled swimming speed of a squirmer, $U/(B_1/2)$, as a function of the squirmer radius for several values of parameter $\varepsilon$. As in \citet{Camley2013}, for a selected dependence of $\varepsilon$ on $\delta$ (e.g. $\varepsilon = \delta/6$), we calculated the swimming speed for a range of $\delta$'s and extrapolated it to $\delta\to 0$.  As can be seen in figure~\ref{fig:epsilon_dependence}a, the regularization parameter $\varepsilon=\delta/12$ gives the swimming velocity that is close to the known value of $B_1/2$ in the limiting case of a pure 2D membrane (membrane in vacuum). Our calculations also show that the scaled swimming velocity $U/(B_1/2)$ is independent of the ratio $B_2/B_1$.

In figure~\ref{fig:epsilon_dependence}b we compare the results of calculations for the swimming speed obtained within the LRT and the BEM for $\varepsilon=\delta/12$. For the direct BEM calculation of the swimming speed and the force distribution we solved simultaneously the equations that impose  no-slip boundary conditions and a zero net force on the swimmer,
\begin{eqnarray}
\label{eq:interiorV}
 U_ \alpha& = &  \sum_{\rm interior\ blobs} \alpha_{\alpha\beta}({\bld x}_j - {\bld x}_i) f_\beta({\bld x}_i),\\
 \label{eq:rimV}
U_\alpha + u_{S\alpha}({\bld x}_j) &=& \sum_{\rm blobs\ on\ rim}  \alpha_{\alpha\beta}({\bld x}_j - {\bld x}_i) f_\beta({\bld x}_i), \\
0&=&\sum_{i=1}^N {\bld f}({\bld x}_i).
\end{eqnarray}
As can be observed in figure~\ref{fig:epsilon_dependence}b, the squirmer swimming velocity decreases with an increase of $a/\ell_S$ ratio. Larger values of $a/\ell_S$ correspond to a larger viscosity of the fluid embedding the membrane, which leads to a stronger traction on the `back' and `belly' of the critter.

\section{Conclusion}
\label{Sec:Conclusion}
We have studied analytically and computationally the locomotion of microscopic organisms confined to a plane of a thin fluid membrane embedded in a bulk fluid of different viscosity. 
In our model the membrane is sufficiently thin, with material particles moving only in the plane of the membrane (the motion in the perpendicular direction is forbidden).

The presence of the bulk fluid allows the  introduction of a hydrodynamic length scale, the Saffman length, that controls the energy exchange between the membrane and the surrounding fluid. 
 By varying the Saffman length, we make our model continuously vary between a pure 2D system (large Saffman length) and a quasi-2D system (small Saffman length).
 The hydrodynamic flows in the quasi-2D membrane have features of both 3D and 2D hydrodynamics. We show that a flagellated swimmer in a viscous film (Saffman length smaller than swimmer characteristic length scale) swims faster than the same swimmer in a 3D fluid. The speedup comes from the effectively larger perpendicular drag coefficient, which arises from the incompressibility of the membrane. On the other hand, a circular squirmer, whose propulsion mechanism does not employ the local drag anisotropy, slows down for smaller Saffman lengths  (in comparison with the squirmer's radius). 
 
The coupling of the membrane with the bulk fluid makes the problem three-dimensional and quite difficult for analytical treatment. We developed numerical schemes based on the boundary element method and the Lorentz reciprocal theorem. We show how the Lorentz reciprocal theorem can be used to simplify the computation of swimming speed,  especially for swimmers such as the squirmer that do not change shape during a stroke. 
 
 While we considered the minimal models of a flagellated swimmer and of a squirmer, our approach can be generalized to other swimmers' geometries and swimming stokes.

\textbf{Acknowledgements}
We thank M. Moelter for helpful discussions.
This work has been supported by a Cottrell College Science Award (TK) and National Science Foundation Grant No. 1437195 (TRP). (CA) gratefully acknowledges support from a Frost Undergraduate Student Research Award.

\appendix
\section{Infinitely long flagellum in 2D fluid}
\label{appA}
In the 2D limit (membrane in vacuum) the system of equations~(\ref{eq:discrete-equation}) becomes
\begin{equation}
\label{eq:discrete-equation-2D}
\left. \begin{array}{c}
\displaystyle
\bld{u}_S(\bld{x}_i)+ \bld{U} = \frac{1}{4\pi\eta_m}\sum_{j = 1}^N \left( \int_{S_j} \bld{G}^p(\bld{x}_i - \bld{x}^\prime) \mathrm{d} \bld{x}^\prime \right)  \bcdot \bld{f}(\bld{x}_{j}) \\[16pt]
\displaystyle
\sum_{j = 1}^N  \bld{f}(\bld{x}_{j}) = 0,
\end{array} \right\}
\end{equation}
where $\bld{G}^p(\bld{x})$ is a 2D periodic Stokeslet,
\begin{equation}
\label{eq:periodic-stokeslet}
\bld{G}^p(\bld{x} -\bld{x}^\prime) = \sum_{m = -\infty}^\infty -{ \bld{\mathrm I}} \ln(q r_m)  + \frac{\bar{\bld{x}}_{m} \bar{\bld{x}}_{m} }{r_m^2},
\end{equation}
with $\bar{\bld{x}} \equiv \{ \bar{x}, \bar{y}\} = \bld{x} - \bld{x}^\prime $,  $\displaystyle\bar{\bld{x}}_{m} = \{\bar{x}  + m \frac{2\pi}{q}, \bar{y}  \}$ and $r_m=|\bar{\bld{x}}_{m}|$. In equations~(\ref{eq:discrete-equation-2D}) and (\ref{eq:periodic-stokeslet}) all variables are dimensional. 
The periodic Stokeslets can be expressed in closed form  \citep{Sauzade2011, Pozrikidis1987} using the analytic formula for the summation, 
\begin{equation}
A = \sum_{m=-\infty}^\infty  \ln(|q r_m|) = \frac{1}{2} \ln \left[ 2 \cosh(q \bar{y}) - 2 \cos(q \bar x) \right].
\end{equation}
The components of the periodic stokeslet can be found as
\begin{eqnarray}
G_{xx}^p &=& - A - A_y + 1, \label{eq:Gxx}\\
G_{xy}^p &=& (q \bar{y}) A_x,\label{eq:Gxy}\\
G_{yy}^p &=& - A + (q \bar{y}) A_y \label{eq:Gyy},
\end{eqnarray}
where $A_x$, $A_y$ indicate the derivatives of $A$ with respect to $q\bar{x}$ and $q\bar{y}$ respectively.
Similar to our treatment of the diagonal terms in equation~(\ref{eq:diag-term-x}) we eliminate the logarithmic singularity by analytic integration,
\begin{eqnarray}
 \int_{S_i} \bld{G}^p(\bld{x}_i - \bld{x}^\prime) \mathrm{d} \bld{x}^\prime  =  { \bld{\mathrm I}} \, 2 \lim_{\varepsilon \to 0} \int_{\varepsilon}^{\Delta s /2} (1 -\log(q z))  \mathrm{d} z =
 { \bld{\mathrm I}} \Delta s ( 1 - \log(q\Delta s/2).
\end{eqnarray}

\section{Swimming velocity of a squirmer in the 2D limit}
\label{appB}
We consider a tangential squirmer with a prescribed surface velocity of the form
\begin{equation}
\boldsymbol{u}_S(\phi) =  u_S(\phi) \hat{\boldsymbol{\phi}}= (B_1\sin\phi + B_2 \sin (2 \phi))\hat{\boldsymbol{\phi}}, 
\end{equation}
with free parameters $B_1$ and $B_2$.
Since the disturbance velocity has only a $\hat{\boldsymbol{\phi}}$ - component, the RHS of equation~(\ref{eq:Lorentz-for-swimmer}) becomes:
\begin{eqnarray}
\label{eq:oint}
\oint_{S(t)} {\boldsymbol{u}}_S \cdot \mathrm{d} {\boldsymbol{F}}^\prime  &=&  \int_{S_w}  u_S(\phi) \hat{\boldsymbol{\phi}} \cdot  ({\sigma^\prime_{r\phi}} \hat{\boldsymbol{\phi}} )\,\mathrm{d} S_w \nonumber \\
&=& 2 h a \int_0^\pi u_S(\phi) \sigma^\prime_{r\phi}  \mathrm{d} \phi,
\end{eqnarray}
where we took into account  $\mathrm{d} S_w =  h a  \mathrm{d} \phi$, where $h$ is the thickness of the membrane. 
Therefore, equation~(\ref{eq:Lorentz-for-swimmer}) becomes 
\begin{equation}
\label{eq:Lorentz-tangential-squirmer}
{\boldsymbol{F}}^\prime(t) \cdot \boldsymbol{U}(t) = - 2 h a \int_0^\pi u_S(\phi) \sigma^\prime_{r\phi} \mathrm{d} \phi.
\end{equation}
The membrane stress tensor element $\sigma_{r\phi}^\prime(r,\phi)$ is determined as
\begin{equation}
 \sigma_{r\phi}^\prime(r,\phi) = - \eta\left[ \frac{1}{r} \frac{\partial {u}^\prime_r}{\partial \phi} + \frac{\partial {u}^\prime_\phi }{\partial r}   - \frac{{u}^\prime_\phi}{r}  \right]
\end{equation}
For a special case of  $a\ll \ell_S$ \citet{Saffman1976} found 
\begin{eqnarray}
{\boldsymbol{F}}^\prime &=& \frac{4 \pi \eta h {\boldsymbol U}^\prime}{ \log(2 \ell_S/a) - \gamma }, \label{eq:force} \\
\sigma_{r\phi}^\prime(r,\phi) &=& \eta \frac{4\alpha \sin\phi}{r^3}, \label{eq:sigma}
\end{eqnarray}
with
\begin{eqnarray}
\label{eq:alpha}
\alpha &=& \frac{a^2 {U}^\prime }{2 ( \gamma - \log(2 \ell_S/a))}. 
\end{eqnarray}
Here $\gamma = 0.577$ is the Euler constant. 

Plugging equations~(\ref{eq:force}), (\ref{eq:sigma}) and (\ref{eq:alpha}) in equation~(\ref{eq:Lorentz-for-swimmer}), after some simplifications we arrive at the squirmer swimming velocity in the limit of $a/\ell_S \ll 1$:
\begin{eqnarray}
U = \int_0^\pi u_S(\phi) \sin\phi  \mathrm{d} \phi= \frac{B_1}{2}.
\end{eqnarray}

\bibliographystyle{jfm}
\bibliography{swimming_references}

\end{document}